\documentclass[letterpaper]{article} 
\usepackage{aaai23}  
\usepackage{times}  
\usepackage{helvet}  
\usepackage{courier}  
\usepackage[hyphens]{url}  
\usepackage{graphicx} 
\usepackage{natbib}  
\usepackage{caption} 
\usepackage{algorithm}
\usepackage{algorithmic}
\usepackage{comment}

\usepackage{newfloat}
\usepackage{listings}
\DeclareCaptionStyle{ruled}{labelfont=normalfont,labelsep=colon,strut=off} 
\floatstyle{ruled}
\newfloat{listing}{tb}{lst}{}
\floatname{listing}{Listing}
\title{Agent Assessment of Others Through the Lens of Self---A Position Paper}

\author{
    Jasmine A. Berry 
}
\affiliations{
    University of Michigan\\

    Robotics Institute\\
    Ann Arbor, Michigan 48109 USA\\
    jasab@umich.edu
}

\usepackage{bibentry}

\begin{document}

\maketitle

\begin{abstract}
The maturation of cognition, from introspection to understanding others, has long been a hallmark of human development. 
This position paper posits that for AI systems to truly emulate or approach human-like interactions, especially within multifaceted environments populated with diverse agents, they must first achieve an in-depth and nuanced understanding of self. 
Drawing parallels with the human developmental trajectory from self-awareness to mentalizing (also called theory of mind), the paper argues that the quality of an autonomous agent's introspective capabilities of self are crucial in mirroring quality human-like understandings of other agents. 
While counterarguments emphasize practicality, computational efficiency, and ethical concerns, this position proposes a development approach, blending algorithmic considerations of self-referential processing.
Ultimately, the vision set forth is not merely of machines that compute but of entities that introspect, empathize, and understand, harmonizing with the complex compositions of human cognition.
\end{abstract}

\section{Introduction}
The ever-expanding frontiers of human-robot interaction (HRI) and artificial intelligence (AI) have introduced a myriad of challenges and opportunities for enabling sophisticated team dynamics. 
Among these challenges is the necessity for agent teaming in mixed-motive situations, which are scenarios where team members (either human or robot) may have partially conflicting goals or objectives. 
While traditional teaming environments typically rely on shared goals for collaboration, mixed-motive situations introduce a dimension of complexity where agents must negotiate, strategize, and sometimes compete, while still ensuring that the overall team objectives are achieved.

The importance of effective agent teaming in such situations cannot be understated. 
As robots and AI systems become more prevalent in human environments – from healthcare to defense, and from manufacturing to domestic settings – there will be numerous situations where human and robotic agents need to cooperate, despite having individual preferences or tasks that might not perfectly align.

Let's address whether an agent’s assessment of the intent and proficiency of other agents \emph{first} relies on its assessment of the intent and proficiency of self and how the notion might correspond with the challenges of agent teaming in mixed-motive situations. 
In such scenarios, each agent's understanding of their own objectives and abilities serves as a foundation for comprehending the perspectives of other agents. 
They might draw analogies from their own motivations and skills when trying to interpret why other agents are behaving in certain ways or making specific decisions.
Successfully navigating mixed-motive situations requires agents to strike a balance between using their own self-assessment as a reference point and being open to understanding and accommodating the diverse intentions and proficiencies of their fellow agents. This alignment of self-assessment and understanding others is a key factor in enabling effective teamwork in situations where agents have differing goals, incentives, and decision-making processes.

\section{Statement of Position}
The indivisible interconnection between an agent's self-perception (i.e., self-awareness, self-consciousness) and its capability to assess others represents a foundational cornerstone in the development and operation of AI systems that are advancing at a pace that often exceeds our progress in understanding their potential.
Drawing from human cognitive processes, the sequence of gaining self-awareness followed by the development of theory of mind—as evidenced in human developmental psychology \cite{baron1991precursors, morin2011self}—underscores the argument that the quality of an agent's assessment of others is deeply rooted in, and perhaps inextricably linked to, the quality of the agent's own self-perception. 
Autonomous agents, like humans, must achieve a sufficient level of self-understanding or model of self \citep{berry2020sensory} as a preliminary step before it can effectively employ mechanisms similar to humans theory of mind (ToM). 
Without a robust self-perception, the agent's ability to understand, anticipate, or interact meaningfully with other entities, be they human or machine, may be significantly compromised. 
Thus, our position postulates that for systems to genuinely excel in complex, human-centric environments—where understanding the intentions, motivations, and cognitive processes of others is vital—they must first be grounded in a comprehensive self-understanding. 

\section{Supporting Arguments}
The progression from self-awareness to theory of mind in developmental psychology offers a persuasive template for understanding agent interactions. 
Without an intuitive understanding of oneself, any attempt to decode others' intentions or emotions would be haphazard, akin to shooting in the dark without a reference point.
\subsection{First Assertion: Philosophical Context}
Philosophers, dating back to the times of Socrates with his edict,``Know Thyself,'' have long stressed the importance of self-awareness as a prerequisite for understanding others \cite{green2017know}. The very nature of introspection, reflection, and existential questioning has its roots in recognizing one's own existence, beliefs, and motivations. 
It is through this lens of self-understanding that one is then able to extrapolate and make educated assumptions or inferences about others. 
As Descartes proclaimed, "I think, therefore I am,"  any agent, to be truly effective in understanding others, must first grapple with its own existential and cognitive architecture.
\subsection{Second Assertion: Psychological \& Behavioral Context}
Psychological studies underscore the evolutionary and developmental importance of self-awareness preceding the development of theory of mind. 
From the `mirror test' in toddlers to behavioral studies in primates, self-recognition often comes before the recognition of others as separate entities with their own thoughts and emotions. 
If this logic extends to artificial agents, the development of their cognitive capabilities should echo the same. 

\subsection{Third Assertion: Robotics Context}
 In the realm of robotics, the challenge is not just theoretical but also practical. 
 For instance, a robot (particularly designed for HRI) needs to know its physical dimensions and capabilities before it can assist or cooperate with a human in a shared task. 
 Only then can it predict and understand human actions and intentions. 
 Furthermore, as robots engage in more sophisticated tasks, like caregiving or companionship, their depth of self-perception directly impacts their ability to understand human goals and intentions.

\section{Counterarguments and Rebuttals}
I acknowledge there is a growing body of research showcasing agents ability to perform successful inference of others' motivations and intentions with emphasis on self-perception \cite{butterfield2009modeling, nemlekar2023transfer, chen2021visual, kosinski2023theory}, which can be the basis of counterarguments to our main argument.

\textbf{Human Development Isn't a Template}: 
Given the distinct differences in biological evolution and engineered design, agents might find alternate paths to understanding and predicting the behaviors of other agents without first achieving a profound sense of self-understanding.

\textbf{Efficiency and Practicality}: Achieving high-level self-awareness in AI systems might be computationally expensive and time-consuming. 
From a practical perspective, it might be more efficient to directly program AI systems with algorithms that predict and understand the parameters of external agents' behaviors without deep introspection.

\textbf{Ethical Implications}: Making machines self-aware might unintentionally lead to ethical dilemmas surrounding machine rights, consciousness, and potential suffering. It's possible to design machines that are effective in human-AI interactions without delving into these morally grey areas \cite{bryson2010robots}.

\textbf{Difference in Learning Mechanisms}: Humans rely on years of experience and social interactions to develop theory of mind. AI, with its ability to process vast amounts of data quickly, might not need to undergo the same introspective processes as humans to achieve a similar outcome in perceiving other agents in mixed-motive environments.

\section{Proposed Solution}
Given the complexity of the subject and the valid points raised in both the argument and counterargument sections, the solution should aim to strike a fair balance.
\begin{enumerate}
\item \textbf{Hybrid Development Model}: Develop AI systems using a hybrid model that combines introspective self-awareness processes with direct external observation. 
By doing so, AI can quickly adapt to external agents and environments while maintaining a degree of self-awareness.
\textbf{Action Steps}: 
Periodically allow AI systems to analyze their actions and decisions, refining their understanding of both self and others.
\item \textbf{Ethical Framework for AI Self-awareness}: Establish a comprehensive ethical framework that addresses the implications of AI self-awareness.
\textbf{Action Steps}: Collaborate with ethicists, AI developers, and public citizens to draft guidelines.
Set boundaries on the depth of self-awareness and self-modification capabilities in AI systems to prevent unforeseen consequences.
\item \textbf{Feedback Mechanisms}: Implement feedback mechanisms that allow software to learn from its environment and refine its self-understanding without necessarily diving deep into introspection.
\textbf{Action Steps}: Embed real-time feedback loops where AI can adjust its actions based on immediate results.
Employ supervised learning where human experts guide an agent's understanding of complex scenarios.
\item \textbf{Dynamic Learning Pathways}: Recognize that not all AI and robotic systems need the same level of self-awareness. Create dynamic learning pathways tailored to the specific needs and roles of individual AI systems.
\textbf{Action Steps}:
For task-specific AI, prioritize efficiency and direct external learning of others.

\end{enumerate}

\section{Conclusion}
The main argument of this work is that agents in mixed-motive situations, akin to humans, must achieve a high quality and nuanced self-perception as a foundational step before it can effectively implement mechanisms similar to the human 'theory of mind'.
Emphasizing the need for self-awareness as a precursor to understanding external entities, this proposal draws parallels from developmental psychology, philosophical principles, and practical implications in multi-agent environments to substantiate the claim that the road map of an agent's cognitive development towards possessing a theory of mind is firmly rooted in its own self-awareness.
While counterarguments rightly point towards the challenges of efficiency and practicality, they too underline an undeniable fact: the quest for self-perception (and even conscious) systems isn't merely a philosophical aspiration but a tangible imperative to ensure responsible and ethical development.

\appendix

\bibliography{aaai23}

\end{document}